\newif\ifforLAUR\forLAURfalse
\ifx\texorpdfstring\undefined\newcommand\texorpdfstring[2]{#1}\fi
\documentclass[aps,prl,twocolumn,superscriptaddress,showpacs,nofootinbib,preprintnumbers,10pt,floatfix]{revtex4-2}

\usepackage{graphicx}
\usepackage[dvipsnames]{xcolor}
\usepackage{amsmath,amsthm,amssymb}
\usepackage{comment}
\usepackage{qcircuit}
\usepackage{tikz}
\usepackage{hyperref}

\makeatletter
\newcommand\showtitleinbib{{\escapechar=`\\ \immediate\write\@auxout{%
\csname citation{REVTEX42Control}\endcsname^^J%
\csname citation{apsrev42Control}\endcsname
}}}
\makeatother

\ifforLAUR\newcommand\mynewcommand[2]{\newcommand#1{}}\else\let\mynewcommand\newcommand\fi

\mynewcommand{\todo}{\colorbox{pink}{\textsc{Todo}}}
\mynewcommand{\scott}{\colorbox{NavyBlue}{\textsc{Scott}}}
\mynewcommand{\junsik}{\colorbox{green}{\textsc{Jun-Sik}}}
\mynewcommand{\tanmoy}{\colorbox{Cyan}{\textsc{Tanmoy}}}

\begin{document}

\title{Control variates for lattice field theory}
\preprint{LA-UR-23-28228}

\author{Tanmoy Bhattacharya}
\email{tanmoy@lanl.gov}
\affiliation{Group T-2, Los Alamos National Laboratory, Los Alamos, NM 87545}
\author{Scott Lawrence}
\email{scott.lawrence-1@colorado.edu}
\affiliation{Department of Physics, University of Colorado, Boulder, CO 80309, USA}
\author{Jun-Sik Yoo}
\email{junsik@lanl.gov}
\affiliation{Group T-2, Los Alamos National Laboratory, Los Alamos, NM 87545}

\date{\today}

\begin{abstract}
In most lattice field theories, correlators are plagued by a signal-to-noise problem of exponential difficulty in the time separation. We propose a method for improving the signal-to-noise ratio, in which control variates are systematically constructed from lattice Schwinger-Dyson relations. The method is demonstrated on various two-dimensional lattices in scalar field theory, and a strategy for scaling to larger systems is explored.
\end{abstract}

\maketitle

With the advent of supercomputing, numerical lattice field theory has become the method of choice for first-principles calculations beyond perturbation theory. In particular, lattice QCD calculations provide the most reliable estimates for quantities of interest in hadronic physics~\cite{USQCD:2022mmc}. It is a systematically improvable scheme in which the only uncertainties stem from the extrapolations to the chiral-continuum-infinite-volume limit and from statistical noise.  For situations where the measure of the path integral---determined by the Euclidean action of the theory---is real and positive, it has become routine to carry out high-statistics, large-volume simulations of lattice  QCD with physical quark masses at a number of lattice spacings, leading to reliable extrapolations.  This is especially true for pseudoscalar meson correlators, whose signal-to-noise ratio stays constant at long Euclidean time separation.

Extending these successes to precise calculations with other mesons and baryons is, however, meeting barriers. In particular, since the interpolating fields used in an interacting field theory couple to multiple asymptotic states, calculations from short separations in Euclidean time have irreducible systematic uncertainties due to contributions involving excited states. For most of these states, at the Euclidean distances needed to isolate the lowest state of interest, the signal-to-noise degrades and the statistical noise becomes insurmountable with today's resources~\cite{Jang:2019vkm}. A canonical example of the signal-to-noise problem is in the baryon-baryon correlator of QCD. An argument of Parisi and Lepage~\cite{Parisi:1983ae,Lepage:1989hd} shows that the signal and noise both fall exponentially, but the exponential fall of the signal is faster than that of the noise, and therefore the signal-to-noise ratio falls exponentially as well. The situation for gluonic correlators is similar: the accuracy with which we can evaluate the expectation values of large Wilson loops or the static-quark potential is severely limited by noise. Since the most accurate determination of the lattice scale is from such gluonic observables~\cite{FlavourLatticeAveragingGroupFLAG:2021npn}, this is a serious problem for any precise lattice calculations~\cite[as an example]{Aubin:2019usy}. In this work we will examine the simpler situation of a scalar field theory in two spacetime dimensions, where the signal falls exponentially while the noise does not; though the methods developed will be generic and expected to be generalizable to other theories.

Previous work has addressed this problem and developed techniques for variance reduction. Of particular interest is a class of techniques in which the obervable of interest can be approximated by an observable that is less expensive to evaluate but whose fluctuations are highly correlated~\citep{Blum:2012uh,Yoon:2018krb}. The statistical fluctuations in both the approximant and the difference can then be evaluated with smaller effort, leading to an improvement in the signal-to-noise performance. A different class of methods starts from the observation that the signal-to-noise problem is closely connected to the sign problem, and the same methods have often been applied to both. In particular, the method of contour deformations used in~\cite{Detmold:2020ncp,Kanwar:2023otc} has its history in approaches to the sign problem; see~\cite{Alexandru:2020wrj} for a review. It relies on the fact that the path integrals defining the expectation values of the observables are analytic functions, as such a continuation of the domain of the path integral to complex field space can keep the value of the integral unchanged. The statistical variance, however, is not analytic, and it can be reduced. 

In this work, we take an alternative course to variance reduction that merges the strengths of the two methods. It is similar to the correlated approximant method in that we find a correction to the integrand whose fluctuations are correlated with the observable.  But, like the contour deformation method, the subtraction is of a term that can be guaranteed to integrate to zero. This method has previously been explored for the the fermion sign problem, via perturbative expansion~\cite{Lawrence:2020kyw} and machine learning~\cite{Lawrence:2022dba}.

Thus, the core of the approach is the construction of a \emph{control variate}: a function of the fields which is correlated with the observable of interest, but can be proven to have vanishing expectation value. Naming the control variate $f$, we define an improved observable $\tilde{\mathcal O} \equiv \mathcal O - f$. First note that since $\langle f \rangle = 0$, this new observable obeys $\langle \tilde{\mathcal O}\rangle = \langle \mathcal O \rangle$. The variance of $\tilde{\mathcal O}$, however, has changed:
\begin{equation}\label{eq:cv-var}
    \langle \tilde{\mathcal O}^2 \rangle - \langle \tilde{\mathcal O}\rangle^2
    = \langle \mathcal O^2\rangle - \langle \mathcal O\rangle^2 - 2 \langle \mathcal O f\rangle + \langle f^2\rangle\text.
\end{equation}
It is apparent that, when the correlation $\langle \mathcal O f\rangle$ is large and positive, the variance of the improved observable is substantially reduced.

It remains to minimize the variance $\langle \tilde{\mathcal O}\rangle$ over a set of functions $f$ that can be guaranteed to integrate to $0$. The main approach we put forward is to obtain a large basis $\{F_i\}$ of functions such that $\langle F_i\rangle = 0$, and then optimize the coefficients $c_i$ that define the control variate $f \equiv \sum_i c_i f_i$. This optimization can be performed efficiently as follows. Measure the covariance matrix of the $F_i$, and the cross-correlations between $F_i$ and the target observable $\mathcal O$, defining
\begin{equation}\label{eq:cv-correlations}
    M_{ij} = \langle F_i F_j\rangle \text{ and } v_i = \langle F_i \mathcal O\rangle\text.
\end{equation}
The optimal control variate is defined to be the linear combination of $F_i$ that minimizes the variance of Eq.~(\ref{eq:cv-var}). Assuming that $M$ and $v$ have been measured accurately, this is readily shown to be given by coefficients
\begin{equation}\label{eq:cv-optimum}
    c = M^{-1} v\text.
\end{equation}
To obtain a basis \(\{F_i\}\), we note that
for any function $g$ of fields the integral of $\partial g$ (where ``$\partial$'' indicates any derivative with respect to the fields) reduces to boundary terms, and therefore $0$ under mild assumptions about the asymptotic behavior\footnote{For the example of scalar field theory used in this paper, $g$ can be constructed from polynomials in the fields as explained below; in the context of compact gauge theories there is no boundary and therefore no restriction on $g$.}.

Having obtained these coefficients and therefore the definition of the control variate $f$, we may compute the expectation value of $\tilde{\mathcal O}$ (and therefore $\langle \mathcal O \rangle$) by standard Monte Carlo methods. It is important, to avoid introducing a bias, that this final computation of the expectation value is performed over a set of samples uncorrelated with those used to determine the coefficients $c$ in the first place.



Let us see how this method works in scalar field theory in two spacetime dimensions. This model is defined by the action
\begin{equation}
    S = \sum_{\langle r ,r'\rangle} \frac {\left(\phi(r) - \phi(r')\right)^2}{2}
    + \sum_r \left[\frac{m^2}{2} \phi(r)^2 + \frac{\lambda}{24!} \phi(r)^4\right]
    \text,
\end{equation}
where $m^2$ (not necessarily non-negative) defines the bare mass and $\lambda$ the bare coupling constant. The first summation is taken over pairs of neighboring lattice sites. Lattice units of $a=1$ are assumed throughout, and we note that the action has a \(\mathbb Z_2\) symmetry \(\phi\to-\phi\).

The object of primary interest for us is the correlation function in the $\mathbb Z_2$-vector channel, defined by
\begin{equation}
    C(\tau) = \sum_x \langle \phi(\tau,x)\phi(0,0)\rangle\text.
\end{equation}
As examined in~\cite{Detmold:2020ncp}, this correlator has a signal-to-noise problem that grows exponentially at large values of $\tau$. We will reduce this signal-to-noise problem by identifying a set of observables that have vanishing expectation values, and constructing a linear combination that is maximally covariant with $\phi(\tau,x)\phi(0,0)$.

To obtain the basis vectors \(\{F_i\}\), we note that for any polynomial $P(\phi)$ of the fields and any choice of site $x$, we have
\begin{equation}
    \Big\langle\frac{\partial P(\phi)}{\partial \phi_x} \Big\rangle = \Big\langle P(\phi) \frac{\partial S(\phi)}{\partial \phi_x}\Big\rangle\text.
    \label{eq:scalar-sd}
\end{equation}
As mentioned above, this identity follows straightforwardly from the fact that the integral of a total derivative reduces to a sum of boundary terms, all of which vanish as long as $P(\phi)$ is sub-exponential. This identity is the lattice version of the Schwinger-Dyson equations described in~\cite{Peskin:1995ev}\footnote{More elaborate identities can be derived; we have deliberately taken the simplest construction.}.

We will restrict ourselves to low-order polynomials. Due to the $\mathbb Z_2$ symmetry of the action, Schwinger-Dyson relations derived from even-order polynomials are uninteresting, relating only $\mathbb Z_2$-odd expectation values which are known to vanish by symmetry. First- and third-order monomials at site $y$ yield the following relations for all pairs of sites $x,y$:
\begin{subequations}
    \begin{eqnarray}
        \delta_{x,y} &=& \Big\langle \phi_y \frac{\partial S(\phi)}{\partial \phi_x}\Big\rangle \label{eq:scalar-sd-1}\\
        3\langle \phi_y^2 \rangle \delta_{x,y} &=& \Big\langle \phi_y^3 \frac{\partial S(\phi)}{\partial \phi_x}\Big\rangle\label{eq:scalar-sd-3}
    \end{eqnarray}
\end{subequations}
Note that, in the case of $\lambda = 0$, the lattice Schwinger-Dyson relations Eq.~\ref{eq:scalar-sd-1} are closed. In particular, they entirely determine the correlation function $C(\tau)$ being studied. A minimization of the variance in this case would be tantamount to solving the theory exactly, reducing the statistical noise to zero. At any nonzero coupling, this is no longer true.  Even the full infinite set of Schwinger-Dyson relations is no longer sufficient to uniquely fix expectation values. This can be seen by noticing that Eq.~(\ref{eq:scalar-sd}) is satisfied on any complex integration contour, as long as the integral of $e^{-S}$ along that contour converges and is non-zero. Multiple non-homologous integration contours are available in the presence of a $\phi^4$ interaction, and therefore the solution to the Schwinger-Dyson relations is not unique.


\begin{figure}
\includegraphics[width=0.95\linewidth]{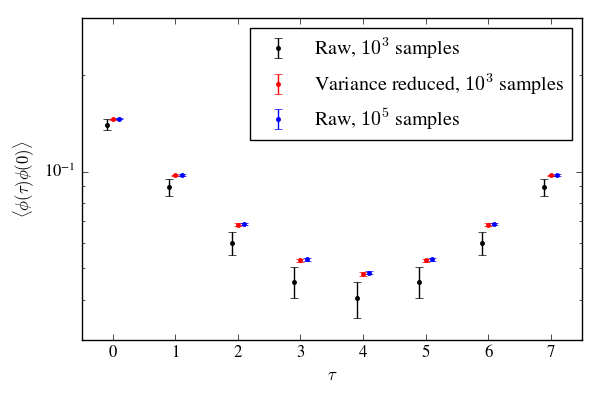}
    \caption{Demonstration of variance reduction via Schwinger-Dyson relations on an $8\times 8$ lattice at $m^2 = 0.1$ and $\lambda = 0.5$. The variance-reduced data is compared to raw correlators with $100$ times as many samples as a check of the correctness of the procedure. Note that the same set of samples is re-used between different values of $\tau$, and therefore the displayed one-sigma uncertainties are correlated.\label{fig:scalar-small}}
\end{figure}

Concretely, our procedure for constructing a control variate from the Schwinger-Dyson relations Eq.~(\ref{eq:scalar-sd-1}) and (\ref{eq:scalar-sd-3}) is as follows. Supposed we have $K$ samples on an $L \times L$ lattice, and the observable of interest is $\mathcal O$. We divide the samples into a training and an evaluation set; here we choose them to be of equal size $\frac K 2$. The training set is used in determining the coefficients defining the control variate, and the evaluation set for computing expectation values\footnote{The purpose of this division into two sets is to avoid biasing the final measurement; if the full set of samples is used for both training and evaluation, final measurements are typically biased towards $0$.}. From Eqs.~(\ref{eq:scalar-sd-1}) and (\ref{eq:scalar-sd-3}), we define the following zero-mean functions, a linear combination of which will become control variates:
\begin{subequations}
    \begin{eqnarray}
        F^{(1)}_{xy} &=& \delta_{xy} - \phi_x \frac{\partial S}{\partial \phi_y}\\
        F^{(3)}_{xy} &=& 3 \delta_{xy}\phi_x^2 - \phi_x^3 \frac{\partial S}{\partial \phi_y}
    \end{eqnarray}
\end{subequations}
The first set of observables will be referred to throughout as first-order, and the second set as the third-order candidates. For convenience, we will denote all the candidate basis functions being used in the optimization as $F_i$ for $i = 1,\ldots,B$.

We now compute, over the training set, the covariance matrix $M$ of the candidates and the correlation $v$ of the candidates with the target observable, as defined by Eq.~(\ref{eq:cv-correlations}). With this done, Eq.~(\ref{eq:cv-optimum}) defines the coefficients of the optimal coefficients $c_i$ defining the control variate, and the variance-reduced unbiased observable
\begin{equation}
    \tilde{\mathcal O} \equiv \mathcal O - \sum_i c_i F_i\text.
\end{equation}
Note that this full procedure must be repeated, yielding a separate control variate, for each observable of interest. In particular, when computing a correlator $\langle \phi(z) \phi(0)\rangle$, a separate control variate must be optimized for each possible value of $z$.

In practice the above procedure has high memory consumption. To make it more feasible on large datasets, we make two simplifications. First, only translationally-invariant sums of $F_{xy}$ need to be used, reducing the size of the basis from $\sim L^2$ to $\sim L$. Second, before computing the correlations $M$ and $v$, we group the samples, still maintaining a sufficiently large number of blocks---specifically, $8$ times the size of the basis being used.

Figure~\ref{fig:scalar-small} shows the efficacy of this approach to variance reduction. On an $8 \times 8$ lattice with bare parameters $m^2 = 0.1$ and $\lambda = 0.5$, we obtain $10^5$ decorrelated samples by MCMC techniques. The correlator determined from the first $10^3$ samples is plotted with and without the variance reduction procedure, using only the first-order candidates $F^{(1)}$. To demonstrate that the correlator is accurate, the results without variance reduction from the full set of $10^5$ samples is also plotted. The variance-reduced correlator is in agreement with the high-statistics correlator, indicating the correctness of the calculation.

\begin{figure}
    \includegraphics[width=.95\linewidth]{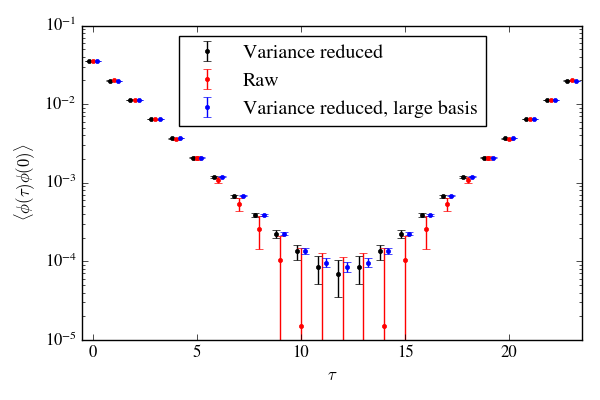}
    \caption{Scalar correlator on a $24 \times 24$ lattice, with bare parameters $m^2 = 0$ and $\lambda = 2$. The raw result---for which the correlator cannot be distinguished from zero at large time separations---is compared to correlators obtained with first-order control variates alone, and those using both first- and third-order control variates. A total of $10^4$ samples were used, with half being reserved for optimization when a control variate is used.\label{fig:scalar-large}}
\end{figure}

To illisutrate the need for higher Schwinger-Dyson equations as we go to more correlated systems, we show in Figure~\ref{fig:scalar-large} the same procedure being performed on a larger, $24 \times 24$ lattice, with bare parameters $m^2 = 0$ and $\lambda = 2$. Variance reduction is performed with $F^{(1)}$ alone and with the combination of $F^{(1)}$ and $F^{(3)}$. At this strong coupling, the use of the additional third-order control variates make a substantial difference at large time separations.

\begin{figure}
    \includegraphics[width=.95\linewidth]{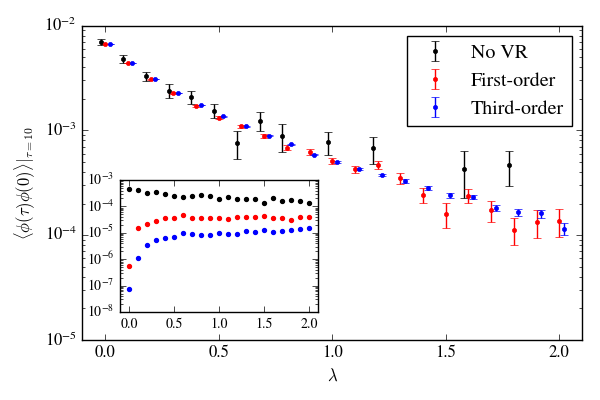}
    \caption{For a fixed time separation of $\tau = 10$, the scalar correlator on a $20 \times 20$ lattice with $m^2 = 0.1$ as a function of the bare coupling. A total of $10^4$ samples were used, with half being reserved for optimization when a control variate is used. Inset is plotted the size of the error bars alone, for the same data.\label{fig:scalar-couplings}}
\end{figure}

This is not surprising. As mentioned above, the first-order Schwinger-Dyson relations of Eq.~(\ref{eq:scalar-sd-1}) give perfect variance elimination at zero coupling, and continue to perform well at weak couplings. At larger couplings, the remaining statistical fluctuations obtained with these relations become larger, and inclusion of higher-order relations (in this work, only Eq.~\ref{eq:scalar-sd-3} was investigated) becomes more important. Figure~\ref{fig:scalar-couplings} shows this in more detail: the estimates obtained as a function of lattice coupling $\lambda$ (on a $20 \times 20$ lattice with bare mass parameters $m^2 = 0.1$) clearly show the need for the third-order term as the coupling gets stronger. At the largest couplings, using the third-order term reduces the number of samples required to get the same size of statistical error by a factor of $\sim 10^2$.

When optimizing the control variate, it is important to have more samples available than degrees of freedom in the control variate. Otherwise, the optimization results in an over-fit to the training data, and the control variate does not generalize to the samples used to compute the expectation value. The control variates discussed so far have $V$ or $2V$ degrees of freedom, and therefore at least that many samples must be used in the optimization procedure.  On the other hand, there is no requirement that the training and evaluation sets need to be of equal size---with limited sampling, it may be advantageous to use a larger training set. Nevertheless, this presents a problem for the scaling of this method to larger systems, where it is common to have far fewer samples available than the volume of the lattice. Thus, many lattice QCD calculations use a \(O(10^3)\) configurations when the number of field degrees of freedom is $\gtrsim\!10^8$~\cite{MILC:2012znn}.  A full treatment of this problem is left for future work, but here we show one approach in the present context.

As a straightforward approach to avoid overfitting, we might try to design a family of control variates with relatively few parameters and optimize over that family. However, that assumes \emph{a priori} knowledge of which basis functions are heavily correlated with the desired observable. In this work we take the less opinionated approach of requiring the vector of coefficients $c$ to be sparse. Each of the $\sim V$ coefficients begins unknown, but we demand that only a small number of them are sizable in the final result.  To learn such a sparse control variate, we introduce an additional term into the objective function:
\begin{equation}\label{eq:l1}
    E_{\mu}^{(p)}(c) = c^T M c - 2 c^T v + \mu \sum_i |c_i|^p\text.
\end{equation}
This is the \(L_p\) regularization, and small positive values of \(p\) enforce high sparsity. The optimization is non-convex for \(p<1\), so we choose to work with the \(L_1\) norm. For well-chosen values of $\mu$, the minimum of this objective will be near a minimum of the original objective function, while having only a small number of contributing entries. This sparsity acts to prevent overfitting. Heuristically, rather than requiring a number of samples greater than the total number of degrees of freedom, we need only to have as many samples as the number of \emph{non-negligible} degrees of freedom.
Of course, this objective function no longer has a closed-form solution for the minimum. Instead, we train the coefficients using the Iterative Shrinkage/Thresholding Algorithm (ISTA)~\cite{beck2009fast}, which performs substantially better than na\"\i{}ve gradient descent for such non-smooth objectives.

Sparsity is not a property of a vector alone, but rather of the components of the vector as written in a certain basis. Equivalently, the $L_1$ norm added to the objective function of Eq.~(\ref{eq:l1}) is not basis-independent. We chose to define the $L_1$ norm with respect to momentum basis; that is, the Fourier transform of the position basis in which $F^{(1)}_{xy}$ was defined above. This approach is motivated by the limit of weak coupling, as at $\lambda=0$ only a single component need be non-zero to control the fluctuations.

Putting this together, we define a control variate
\begin{equation}
    f = \frac 1 {L^2} \sum_{p,q} \tilde c_{p,q} \sum_{x,y} e^{i p x + i q y} F_{xy}^{(1)}
\end{equation}
where $p,q$ are summed over all momentum modes on the lattice while $x,y$ are summed over all $L^2$ lattice sites---where, as before, we have used translational invariance to reduce the implemented basis size. We optimize the coefficients $\tilde c_{p,q}$ to minimize the objective function $E_\mu^{(1)}(c)$ defined in Eq.~(\ref{eq:l1}). Different values of $\mu$ must be tried to find a control variate that makes an acceptable tradeoff between sparsity (avoiding overfitting) and quality; there is no guarantee that such a tradeoff is available.

\begin{figure}
\includegraphics[width=.95\linewidth]{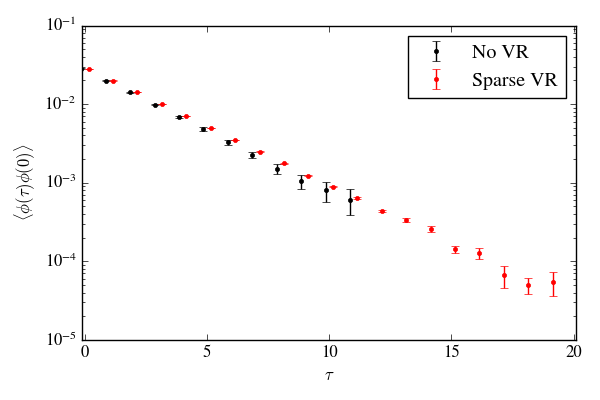}
\caption{Correlator on a $50 \times 50$ lattice with $m^2 = \lambda = 0.1$, trained under $L^1$ regularization with $\mu = 10^{-4}$ as described in the text. A total of $10^3$ samples are used. For clarity, points within $2\sigma$ of $0$ are not plotted.\label{fig:sparse}}
\end{figure}

Figure~\ref{fig:sparse} shows the result of this optimization on a $50 \times 50$ lattice at bare lattice parameters $m^2 = \lambda = 0.1$. Without $L_1$ regularization, at least $2.5 \times 10^4$ samples would be required to avoid overfitting; here only $5 \times 10^2$ samples were used in the optimization of the control variates with an \(L_1\) regularization strength of \(\mu=10^{-4}\), with another $5 \times 10^2$ used to compute the expectation value of the improved observable. Each data point corresponds to a different observable and therefore a different set of coefficients, but the typical vector of coefficients had $\sim 100$ elements of appreciable size, thus allowing optimization with no noticeable overfitting with only 500 samples.

Thus, we have developed a method in which straightforward optimization techniques can be used to choose combinations of Schwinger-Dyson equations to reduce the variance of observables. The strength of the method lies in the fact that existing ensembles of field configurations can be used, the improved observable is exactly equivalent to the original, and the path-integral measure is not distorted alleviating any concerns about overlap of distributions.  Additionally, the use of Schwinger-Dyson equations allows physical intuition to guide the selection of the sparsity penalty that makes the method workable.  The examination of the efficacy of the method for multi-point functions and for non-scalar fields is currently under investigation.

\begin{acknowledgments}
We are indebted to Rajan Gupta, Frederic Koehler, and Boram Yoon for helpful discussions over the course of this work. S.L.~is grateful for the hospitality of the Los Alamos National Laboratory, where this project began. S.L.~is supported by the U.S.~Department of Energy under Contract No.~DE-SC0017905, and T.B.~and J.-S.Y.~under Contract No.~DE-AC52-06NA25396. T.B.~and J.-S.Y.~were also supported by the LDRD program of the Los Alamos National Laboratory. Code used in the demonstrations in this paper is available at~\cite{gitlab}.\\[\baselineskip]\vrule width0pt
\end{acknowledgments}

\bibliographystyle{apsrev4-2}
\showtitleinbib
\bibliography{refs}

\begin{thebibliography}{17}%
\makeatletter
\providecommand \@ifxundefined [1]{%
 \@ifx{#1\undefined}
}%
\providecommand \@ifnum [1]{%
 \ifnum #1\expandafter \@firstoftwo
 \else \expandafter \@secondoftwo
 \fi
}%
\providecommand \@ifx [1]{%
 \ifx #1\expandafter \@firstoftwo
 \else \expandafter \@secondoftwo
 \fi
}%
\providecommand \natexlab [1]{#1}%
\providecommand \enquote  [1]{``#1''}%
\providecommand \bibnamefont  [1]{#1}%
\providecommand \bibfnamefont [1]{#1}%
\providecommand \citenamefont [1]{#1}%
\providecommand \href@noop [0]{\@secondoftwo}%
\providecommand \href [0]{\begingroup \@sanitize@url \@href}%
\providecommand \@href[1]{\@@startlink{#1}\@@href}%
\providecommand \@@href[1]{\endgroup#1\@@endlink}%
\providecommand \@sanitize@url [0]{\catcode `\\12\catcode `\$12\catcode
  `\&12\catcode `\#12\catcode `\^12\catcode `\_12\catcode `\%12\relax}%
\providecommand \@@startlink[1]{}%
\providecommand \@@endlink[0]{}%
\providecommand \url  [0]{\begingroup\@sanitize@url \@url }%
\providecommand \@url [1]{\endgroup\@href {#1}{\urlprefix }}%
\providecommand \urlprefix  [0]{URL }%
\providecommand \Eprint [0]{\href }%
\providecommand \doibase [0]{https://doi.org/}%
\providecommand \selectlanguage [0]{\@gobble}%
\providecommand \bibinfo  [0]{\@secondoftwo}%
\providecommand \bibfield  [0]{\@secondoftwo}%
\providecommand \translation [1]{[#1]}%
\providecommand \BibitemOpen [0]{}%
\providecommand \bibitemStop [0]{}%
\providecommand \bibitemNoStop [0]{.\EOS\space}%
\providecommand \EOS [0]{\spacefactor3000\relax}%
\providecommand \BibitemShut  [1]{\csname bibitem#1\endcsname}%
\let\auto@bib@innerbib\@empty
\bibitem [{\citenamefont {Kronfeld}\ \emph {et~al.}(2022)\citenamefont
  {Kronfeld} \emph {et~al.}}]{USQCD:2022mmc}%
  \BibitemOpen
  \bibfield  {author} {\bibinfo {author} {\bibfnamefont {A.~S.}\ \bibnamefont
  {Kronfeld}} \emph {et~al.} (\bibinfo {collaboration} {USQCD}),\ }\href@noop
  {} {\bibinfo {title} {{Lattice QCD and Particle Physics}}} (\bibinfo {year}
  {2022}),\ \Eprint {https://arxiv.org/abs/2207.07641} {arXiv:2207.07641
  [hep-lat]} \BibitemShut {NoStop}%
\bibitem [{\citenamefont {Jang}\ \emph {et~al.}(2020)\citenamefont {Jang},
  \citenamefont {Gupta}, \citenamefont {Yoon},\ and\ \citenamefont
  {Bhattacharya}}]{Jang:2019vkm}%
  \BibitemOpen
  \bibfield  {author} {\bibinfo {author} {\bibfnamefont {Y.-C.}\ \bibnamefont
  {Jang}}, \bibinfo {author} {\bibfnamefont {R.}~\bibnamefont {Gupta}},
  \bibinfo {author} {\bibfnamefont {B.}~\bibnamefont {Yoon}},\ and\ \bibinfo
  {author} {\bibfnamefont {T.}~\bibnamefont {Bhattacharya}},\ }\bibfield
  {title} {\bibinfo {title} {{Axial Vector Form Factors from Lattice QCD that
  Satisfy the PCAC Relation}},\ }\href
  {https://doi.org/10.1103/PhysRevLett.124.072002} {\bibfield  {journal}
  {\bibinfo  {journal} {Phys. Rev. Lett.}\ }\textbf {\bibinfo {volume} {124}},\
  \bibinfo {pages} {072002} (\bibinfo {year} {2020})},\ \Eprint
  {https://arxiv.org/abs/1905.06470} {arXiv:1905.06470 [hep-lat]} \BibitemShut
  {NoStop}%
\bibitem [{\citenamefont {Parisi}(1984)}]{Parisi:1983ae}%
  \BibitemOpen
  \bibfield  {author} {\bibinfo {author} {\bibfnamefont {G.}~\bibnamefont
  {Parisi}},\ }\bibfield  {title} {\bibinfo {title} {{The Strategy for
  Computing the Hadronic Mass Spectrum}},\ }\href
  {https://doi.org/10.1016/0370-1573(84)90081-4} {\bibfield  {journal}
  {\bibinfo  {journal} {Phys. Rept.}\ }\textbf {\bibinfo {volume} {103}},\
  \bibinfo {pages} {203} (\bibinfo {year} {1984})}\BibitemShut {NoStop}%
\bibitem [{\citenamefont {Lepage}(1989)}]{Lepage:1989hd}%
  \BibitemOpen
  \bibfield  {author} {\bibinfo {author} {\bibfnamefont {G.~P.}\ \bibnamefont
  {Lepage}},\ }\bibfield  {title} {\bibinfo {title} {{The Analysis of
  Algorithms for Lattice Field Theory}},\ }in\ \href@noop {} {\emph {\bibinfo
  {booktitle} {{Theoretical Advanced Study Institute in Elementary Particle
  Physics}}}}\ (\bibinfo {year} {1989})\BibitemShut {NoStop}%
\bibitem [{\citenamefont {Aoki}\ \emph {et~al.}(2022)\citenamefont {Aoki} \emph
  {et~al.}}]{FlavourLatticeAveragingGroupFLAG:2021npn}%
  \BibitemOpen
  \bibfield  {author} {\bibinfo {author} {\bibfnamefont {Y.}~\bibnamefont
  {Aoki}} \emph {et~al.} (\bibinfo {collaboration} {Flavour Lattice Averaging
  Group (FLAG)}),\ }\bibfield  {title} {\bibinfo {title} {{FLAG Review 2021}},\
  }\href {https://doi.org/10.1140/epjc/s10052-022-10536-1} {\bibfield
  {journal} {\bibinfo  {journal} {Eur. Phys. J. C}\ }\textbf {\bibinfo {volume}
  {82}},\ \bibinfo {pages} {869} (\bibinfo {year} {2022})},\ \Eprint
  {https://arxiv.org/abs/2111.09849} {arXiv:2111.09849 [hep-lat]} \BibitemShut
  {NoStop}%
\bibitem [{\citenamefont {Aubin}\ \emph {et~al.}(2020)\citenamefont {Aubin},
  \citenamefont {Blum}, \citenamefont {Tu}, \citenamefont {Golterman},
  \citenamefont {Jung},\ and\ \citenamefont {Peris}}]{Aubin:2019usy}%
  \BibitemOpen
  \bibfield  {author} {\bibinfo {author} {\bibfnamefont {C.}~\bibnamefont
  {Aubin}}, \bibinfo {author} {\bibfnamefont {T.}~\bibnamefont {Blum}},
  \bibinfo {author} {\bibfnamefont {C.}~\bibnamefont {Tu}}, \bibinfo {author}
  {\bibfnamefont {M.}~\bibnamefont {Golterman}}, \bibinfo {author}
  {\bibfnamefont {C.}~\bibnamefont {Jung}},\ and\ \bibinfo {author}
  {\bibfnamefont {S.}~\bibnamefont {Peris}},\ }\bibfield  {title} {\bibinfo
  {title} {{Light quark vacuum polarization at the physical point and
  contribution to the muon $g-2$}},\ }\href
  {https://doi.org/10.1103/PhysRevD.101.014503} {\bibfield  {journal} {\bibinfo
   {journal} {Phys. Rev. D}\ }\textbf {\bibinfo {volume} {101}},\ \bibinfo
  {pages} {014503} (\bibinfo {year} {2020})},\ \Eprint
  {https://arxiv.org/abs/1905.09307} {arXiv:1905.09307 [hep-lat]} \BibitemShut
  {NoStop}%
\bibitem [{\citenamefont {Blum}\ \emph {et~al.}(2013)\citenamefont {Blum},
  \citenamefont {Izubuchi},\ and\ \citenamefont {Shintani}}]{Blum:2012uh}%
  \BibitemOpen
  \bibfield  {author} {\bibinfo {author} {\bibfnamefont {T.}~\bibnamefont
  {Blum}}, \bibinfo {author} {\bibfnamefont {T.}~\bibnamefont {Izubuchi}},\
  and\ \bibinfo {author} {\bibfnamefont {E.}~\bibnamefont {Shintani}},\
  }\bibfield  {title} {\bibinfo {title} {{New class of variance-reduction
  techniques using lattice symmetries}},\ }\href
  {https://doi.org/10.1103/PhysRevD.88.094503} {\bibfield  {journal} {\bibinfo
  {journal} {Phys. Rev. D}\ }\textbf {\bibinfo {volume} {88}},\ \bibinfo
  {pages} {094503} (\bibinfo {year} {2013})},\ \Eprint
  {https://arxiv.org/abs/1208.4349} {arXiv:1208.4349 [hep-lat]} \BibitemShut
  {NoStop}%
\bibitem [{\citenamefont {Yoon}\ \emph {et~al.}(2019)\citenamefont {Yoon},
  \citenamefont {Bhattacharya},\ and\ \citenamefont {Gupta}}]{Yoon:2018krb}%
  \BibitemOpen
  \bibfield  {author} {\bibinfo {author} {\bibfnamefont {B.}~\bibnamefont
  {Yoon}}, \bibinfo {author} {\bibfnamefont {T.}~\bibnamefont {Bhattacharya}},\
  and\ \bibinfo {author} {\bibfnamefont {R.}~\bibnamefont {Gupta}},\ }\bibfield
   {title} {\bibinfo {title} {{Machine Learning Estimators for Lattice QCD
  Observables}},\ }\href {https://doi.org/10.1103/PhysRevD.100.014504}
  {\bibfield  {journal} {\bibinfo  {journal} {Phys. Rev. D}\ }\textbf {\bibinfo
  {volume} {100}},\ \bibinfo {pages} {014504} (\bibinfo {year} {2019})},\
  \Eprint {https://arxiv.org/abs/1807.05971} {arXiv:1807.05971 [hep-lat]}
  \BibitemShut {NoStop}%
\bibitem [{\citenamefont {Detmold}\ \emph {et~al.}(2020)\citenamefont
  {Detmold}, \citenamefont {Kanwar}, \citenamefont {Wagman},\ and\
  \citenamefont {Warrington}}]{Detmold:2020ncp}%
  \BibitemOpen
  \bibfield  {author} {\bibinfo {author} {\bibfnamefont {W.}~\bibnamefont
  {Detmold}}, \bibinfo {author} {\bibfnamefont {G.}~\bibnamefont {Kanwar}},
  \bibinfo {author} {\bibfnamefont {M.~L.}\ \bibnamefont {Wagman}},\ and\
  \bibinfo {author} {\bibfnamefont {N.~C.}\ \bibnamefont {Warrington}},\
  }\bibfield  {title} {\bibinfo {title} {{Path integral contour deformations
  for noisy observables}},\ }\href
  {https://doi.org/10.1103/PhysRevD.102.014514} {\bibfield  {journal} {\bibinfo
   {journal} {Phys. Rev. D}\ }\textbf {\bibinfo {volume} {102}},\ \bibinfo
  {pages} {014514} (\bibinfo {year} {2020})},\ \Eprint
  {https://arxiv.org/abs/2003.05914} {arXiv:2003.05914 [hep-lat]} \BibitemShut
  {NoStop}%
\bibitem [{\citenamefont {Kanwar}\ \emph {et~al.}(2023)\citenamefont {Kanwar},
  \citenamefont {Lovato}, \citenamefont {Rocco},\ and\ \citenamefont
  {Wagman}}]{Kanwar:2023otc}%
  \BibitemOpen
  \bibfield  {author} {\bibinfo {author} {\bibfnamefont {G.}~\bibnamefont
  {Kanwar}}, \bibinfo {author} {\bibfnamefont {A.}~\bibnamefont {Lovato}},
  \bibinfo {author} {\bibfnamefont {N.}~\bibnamefont {Rocco}},\ and\ \bibinfo
  {author} {\bibfnamefont {M.}~\bibnamefont {Wagman}},\ }\href@noop {}
  {\bibinfo {title} {{Mitigating Green's function Monte Carlo signal-to-noise
  problems using contour deformations}}} (\bibinfo {year} {2023}),\ \Eprint
  {https://arxiv.org/abs/2304.03229} {arXiv:2304.03229 [nucl-th]} \BibitemShut
  {NoStop}%
\bibitem [{\citenamefont {Alexandru}\ \emph {et~al.}(2022)\citenamefont
  {Alexandru}, \citenamefont {Basar}, \citenamefont {Bedaque},\ and\
  \citenamefont {Warrington}}]{Alexandru:2020wrj}%
  \BibitemOpen
  \bibfield  {author} {\bibinfo {author} {\bibfnamefont {A.}~\bibnamefont
  {Alexandru}}, \bibinfo {author} {\bibfnamefont {G.}~\bibnamefont {Basar}},
  \bibinfo {author} {\bibfnamefont {P.~F.}\ \bibnamefont {Bedaque}},\ and\
  \bibinfo {author} {\bibfnamefont {N.~C.}\ \bibnamefont {Warrington}},\
  }\bibfield  {title} {\bibinfo {title} {{Complex paths around the sign
  problem}},\ }\href {https://doi.org/10.1103/RevModPhys.94.015006} {\bibfield
  {journal} {\bibinfo  {journal} {Rev. Mod. Phys.}\ }\textbf {\bibinfo {volume}
  {94}},\ \bibinfo {pages} {015006} (\bibinfo {year} {2022})},\ \Eprint
  {https://arxiv.org/abs/2007.05436} {arXiv:2007.05436 [hep-lat]} \BibitemShut
  {NoStop}%
\bibitem [{\citenamefont {Lawrence}(2020)}]{Lawrence:2020kyw}%
  \BibitemOpen
  \bibfield  {author} {\bibinfo {author} {\bibfnamefont {S.}~\bibnamefont
  {Lawrence}},\ }\bibfield  {title} {\bibinfo {title} {{Perturbative Removal of
  a Sign Problem}},\ }\href {https://doi.org/10.1103/PhysRevD.102.094504}
  {\bibfield  {journal} {\bibinfo  {journal} {Phys. Rev. D}\ }\textbf {\bibinfo
  {volume} {102}},\ \bibinfo {pages} {094504} (\bibinfo {year} {2020})},\
  \Eprint {https://arxiv.org/abs/2009.10901} {arXiv:2009.10901 [hep-lat]}
  \BibitemShut {NoStop}%
\bibitem [{\citenamefont {Lawrence}\ and\ \citenamefont
  {Yamauchi}(2023)}]{Lawrence:2022dba}%
  \BibitemOpen
  \bibfield  {author} {\bibinfo {author} {\bibfnamefont {S.}~\bibnamefont
  {Lawrence}}\ and\ \bibinfo {author} {\bibfnamefont {Y.}~\bibnamefont
  {Yamauchi}},\ }\bibfield  {title} {\bibinfo {title} {{Deep learning of
  fermion sign fluctuations}},\ }\href
  {https://doi.org/10.1103/PhysRevD.107.114505} {\bibfield  {journal} {\bibinfo
   {journal} {Phys. Rev. D}\ }\textbf {\bibinfo {volume} {107}},\ \bibinfo
  {pages} {114505} (\bibinfo {year} {2023})},\ \Eprint
  {https://arxiv.org/abs/2212.14606} {arXiv:2212.14606 [hep-lat]} \BibitemShut
  {NoStop}%
\bibitem [{\citenamefont {Peskin}\ and\ \citenamefont
  {Schroeder}(1995)}]{Peskin:1995ev}%
  \BibitemOpen
  \bibfield  {author} {\bibinfo {author} {\bibfnamefont {M.~E.}\ \bibnamefont
  {Peskin}}\ and\ \bibinfo {author} {\bibfnamefont {D.~V.}\ \bibnamefont
  {Schroeder}},\ }\href@noop {} {\emph {\bibinfo {title} {{An Introduction to
  quantum field theory}}}}\ (\bibinfo  {publisher} {Addison-Wesley},\ \bibinfo
  {address} {Reading, USA},\ \bibinfo {year} {1995})\BibitemShut {NoStop}%
\bibitem [{\citenamefont {Bazavov}\ \emph {et~al.}(2013)\citenamefont {Bazavov}
  \emph {et~al.}}]{MILC:2012znn}%
  \BibitemOpen
  \bibfield  {author} {\bibinfo {author} {\bibfnamefont {A.}~\bibnamefont
  {Bazavov}} \emph {et~al.} (\bibinfo {collaboration} {MILC}),\ }\bibfield
  {title} {\bibinfo {title} {{Lattice QCD Ensembles with Four Flavors of Highly
  Improved Staggered Quarks}},\ }\href
  {https://doi.org/10.1103/PhysRevD.87.054505} {\bibfield  {journal} {\bibinfo
  {journal} {Phys. Rev. D}\ }\textbf {\bibinfo {volume} {87}},\ \bibinfo
  {pages} {054505} (\bibinfo {year} {2013})},\ \Eprint
  {https://arxiv.org/abs/1212.4768} {arXiv:1212.4768 [hep-lat]} \BibitemShut
  {NoStop}%
\bibitem [{\citenamefont {Beck}\ and\ \citenamefont
  {Teboulle}(2009)}]{beck2009fast}%
  \BibitemOpen
  \bibfield  {author} {\bibinfo {author} {\bibfnamefont {A.}~\bibnamefont
  {Beck}}\ and\ \bibinfo {author} {\bibfnamefont {M.}~\bibnamefont
  {Teboulle}},\ }\bibfield  {title} {\bibinfo {title} {A fast iterative
  shrinkage-thresholding algorithm for linear inverse problems},\ }\href
  {https://doi.org/10.1137/080716542} {\bibfield  {journal} {\bibinfo
  {journal} {SIAM journal on imaging sciences}\ }\textbf {\bibinfo {volume}
  {2}},\ \bibinfo {pages} {183} (\bibinfo {year} {2009})}\BibitemShut {NoStop}%
\bibitem [{git(2023)}]{gitlab}%
  \BibitemOpen
  \href {https://gitlab.com/s.lawrence/latticecv} {\bibinfo {title}
  {https://gitlab.com/s.lawrence/latticecv}},\ \bibinfo {howpublished} {(Source
  on Gitlab)} (\bibinfo {year} {2023})\BibitemShut {NoStop}%
\end{thebibliography}%

\end{document}